\documentclass[reprint,prc,aps]{revtex4-1}

\usepackage{graphicx}
\usepackage{dcolumn}
\usepackage{bm}

\begin{document}


\title{
The origin of the background radioactive isotope \\
$^{127}$Xe in the sample of Xe enriched in $^{124}$Xe
}

\author{
Yu.M.~Gavrilyuk$^\dag$, A.M.~Gangapshev$^\dag$, V.V.~Kazalov$^\dag$,
V.V.~Kuzminov$^\dag$,
S.I.~Panasenko$^{\ddag}$,
S.S.~Ratkevich$^{\ddag}$,
D.A.~Tekueva$^\dag$
and S.P.~Yakimenko$^\dag$
}
\affiliation{$^{\dag}$ Baksan Neutrino Observatory INR RAS, Russia}
\affiliation{$^{\ddag}$ V.N.Karazin Kharkiv National University, Ukraine}

\date{\today}%

\begin{abstract}
The results of investigation of $^{127}$Xe radioactive isotope production in the xenon sample enriched in
$^{124}$Xe, $^{126}$Xe, $^{128}$Xe are presented.
The isotope is supposed to be the source of the background events in the low-background experiment on search for $2K$-capture of $^{124}$Xe. In this work we consider two channels of $^{127}$Xe production:
the neutron knock-out from $^{128}$Xe nucleus by cosmogenic muons
and the neutron capture by $^{126}$Xe nucleus.
For the first channel the upper limit of the cross section of $^{127}$Xe production was found to be $\sigma \leq 0.007 \cdot 10^{-24}$~cm$^2$ at 95\% C.L.
For the second channel the value obtained for the cross section was found to be equal to $\sigma =(2.74 \pm 0.4)\cdot 10^{-24}$~cm$^2$, which coincides well, within the statistical error, with reference value.
\end{abstract}
\maketitle

\section{Introduction}

Several experiments to search for rare process like $2\beta2\nu$-decay and its variety in the form of $2K2\nu$-capture are carried out at the Baksan Neutrino Observatory, INR RAS \cite{r1}. Such processes are difficult to detect as there is only low energy characteristic photon that is available for registration (total energy release $\sim (25 \div 100)$ keV). The following nuclei are considered as most promising to search for this process: $^{78}$Kr$\rightarrow ^{78}$Se, $^{96}$Ru$\rightarrow ^{96}$Mo, $^{106}$Cd$\rightarrow ^{106}$Pd, $^{124}$Xe$\rightarrow ^{124}$Te, $^{130}$Ba$\rightarrow ^{130}$Xe, $^{136}$Ce$\rightarrow ^{136}$Ba. Two of these isotopes are gases ($^{78}$Kr, $^{124}$Xe), and their kinetic energy of transition, Q, is the largest in this list. It is easy to develop and construct the detection system, a source-detector, on the basis of these gases.

A large proportional counter (LPC) of high pressure with a casing of M1-grade copper has been used to register the process of $2K$-capture in $^{124}$Xe. The details of the counter construction and the technique of the experiment are described in \cite{r2, r3}. At the first stage of the experiment the sample of 12 L (sample no.~1) of Xe enriched in $^{124}$Xe up to 63.3\%
(44 g) has been used \cite{r4}. At the second stage of experiment the sample of Xe was of 50 L (sample no.~2) and enriched in $^{124}$Xe up to 21\%
(58.6 g). Sample no.~2 was composed of sample no.~1 and of 58 L of Xe (sample no.~3) enriched in $^{124}$Xe up to 7.5\%.

\begin{figure}
\begin{center}
\includegraphics*[width=3.0 in,angle=0.]{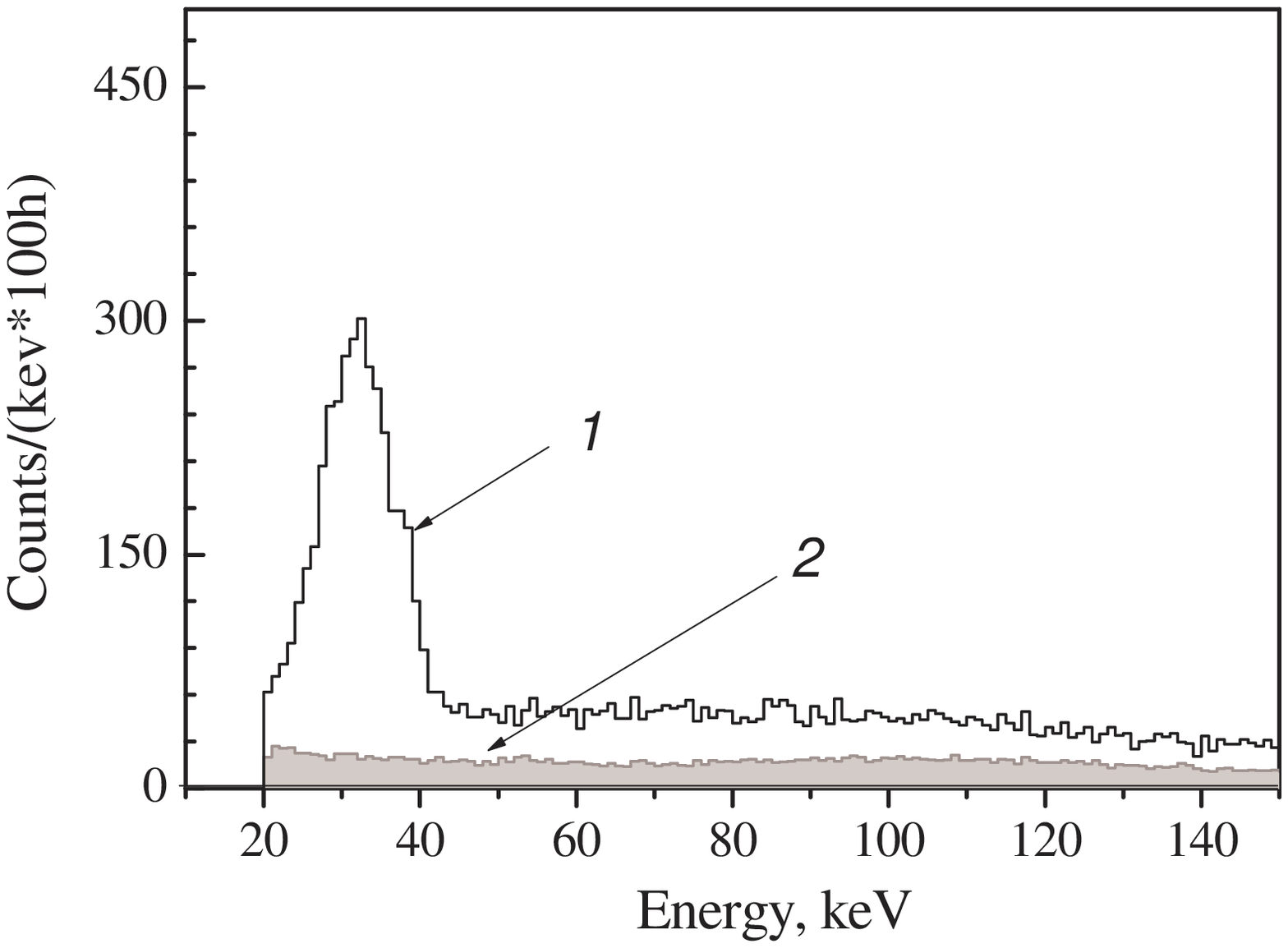}%
\caption{\label{fig1}
Amplitude spectra of LPC (5 atm) with xenon sample no.~3 - spectrum \emph{1},
and radioactive pure xenon - spectrum \emph{2}.
}
\end{center}
\end{figure}

\begin{figure}
\begin{center}
\includegraphics*[width=3.0 in,angle=0.]{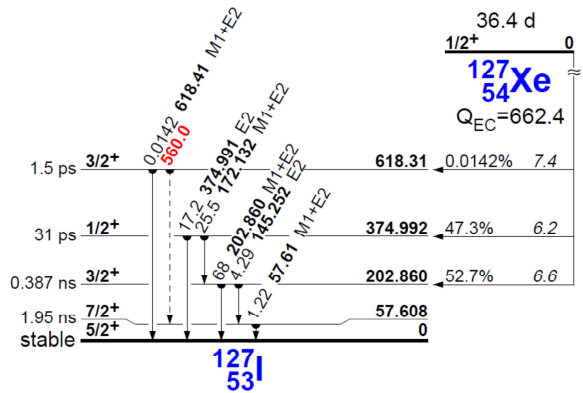}%
\caption{\label{fig2}
The decay scheme of isotope $^{127}$Xe \cite{r5}.}
\end{center}
\end{figure}

At the preparatory stage of the experiment, during three months before the working sample was prepared, several background measurements using 58 L of Xe has been carried out \cite{r4} and a peak from the unknown source has been detected in the $\sim 33$ keV region. (see Fig.\ref{fig1}).
The source of this peak was associated with radioactive isotope $^{127}$Xe which decays
by electron capture (half-life of 36.4 days, Q$_{EC}=662.3$ keV), producing $^{127}$I.
The decay scheme of isotope $^{127}$Xe \cite{r5} is shown in Fig.\ref{fig2}.
$K$-capture yields a 33.2 keV energy release.

After Xe sample no.~3 has been stored under low-background conditions \cite{r6} during three half-lives of $^{127}$Xe, it was used to prepare LPC main gas sample for further measurements. As the nature of $^{127}$Xe presence in sample no.~3 was unknown it was decided to carry out several measurements to make the test of  the cross sections of different channels of its creation.

\section{Method of measurements}

In our work we consider two channels of $^{127}$Xe production. The choice of these channels was determined by the isotopic composition of sample no.~3. Its full isotopic composition is given in Table.

\begin{table*}[pt]
\caption{
Characteristics of samples of xenon
no.~3
and no.~4
}
\begin{tabular}
{|l| c c c c c c c c c|} \hline \hline
 \multicolumn{1}{|c} ~ &  \multicolumn{9}{|c|}{Xenon isotopes}   \\
 \cline{2-10}
 \multicolumn{1}{|c|} {Samples, \emph{L}} & {\small{124}} & {\small{126}} & {\small{128}} & {\small{129}} & {\small{130}} & {\small{131}} & {\small{132}}  & {\small{134}}  & {\small{136}} \\
\cline{2-10}
  ~  &   \multicolumn{9}{c|}{Content, \emph{L} }  \\
\hline

No.~3, {\small{58}} & {\small{4.33}} & {\small{15.24}} & {\small{24.147}}  & {\small{14.033}} &  {\small{0.0511}} & {\small{0.0394}} & {\small{0.0168}} & {\small{0.0606}} & {\small{0.0543}}\\


No.~4, {\small{18}} & {\small{1.31}} & {\small{4.68}} & {\small{7.54}}  & {\small{4.41}} &  {\small{0.0162}} & {\small{0.0126}} & {\small{0.0054}} & {\small{0.0198}} & {\small{0.018}}\\

\hline \hline

\end{tabular}
\end{table*}

As is seen from Table the following two channels of $^{127}$Xe production are most probable:
neutron knock-out from $^{128}$Xe nucleus by cosmogenic muons and the neutron capture by $^{126}$Xe nucleus.
Two independent measurements to test the production of $^{127}$Xe channels were carried out. In these measurements the xenon of 18 L (sample no.~4) was used. It remained after main gas sample (sample no.~2) preparation to search for $2K$-capture of $^{124}$Xe.

\begin{figure}[pt]
\begin{center}
\includegraphics*[width=3.0 in,angle=0.]{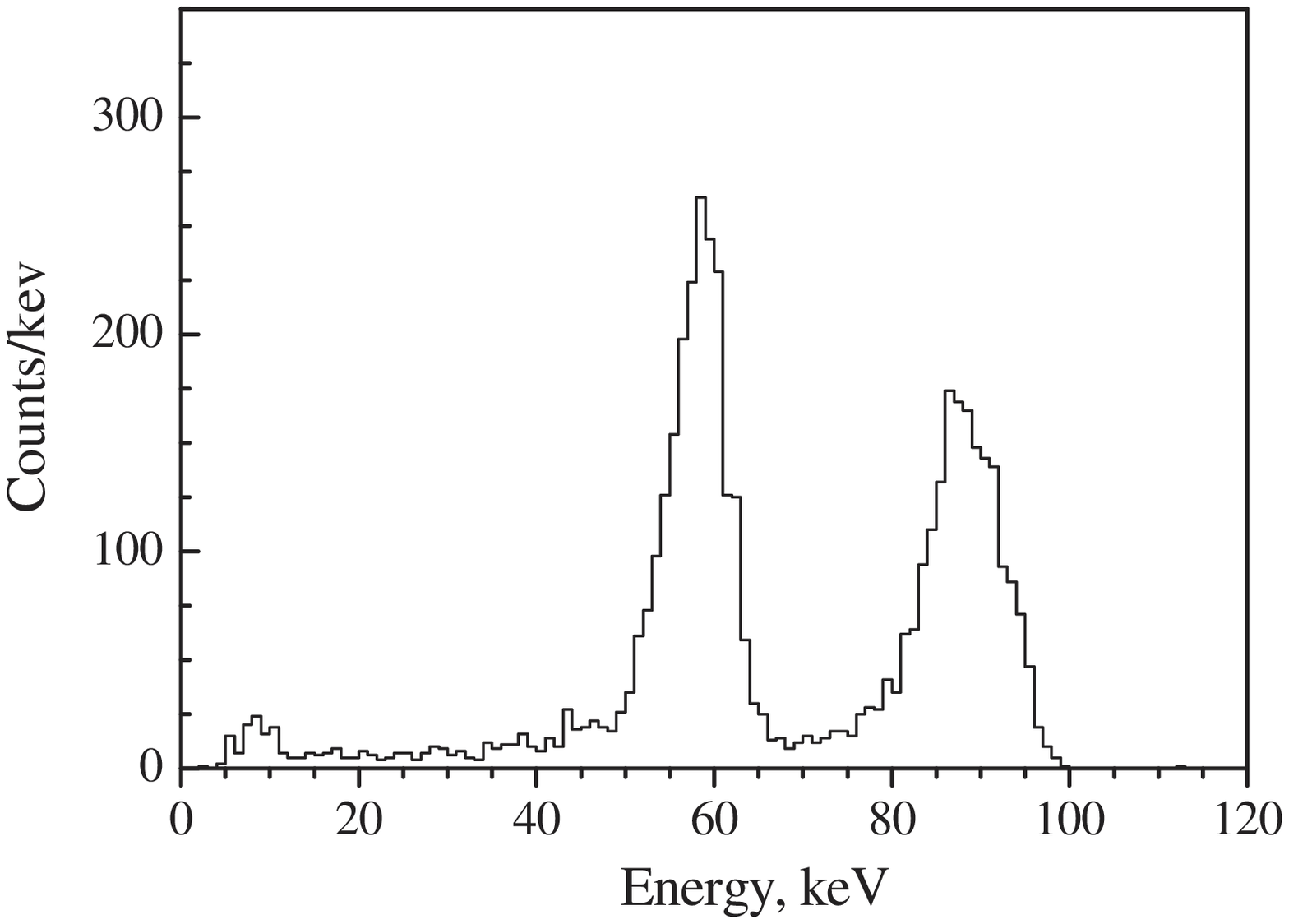}%
\caption{\label{fig3}
Spectrum of $^{109}$Cd source used in calibration of LPC.}
\end{center}
\end{figure}

As mentioned above, the LPC counter was used to investigate the process of $^{127}$Xe production, and $^{109}$Cd source was used for its calibration. The source has two gamma lines, 22 keV and 88 keV, but as the wall of LPC is quite thick (6.5 mm) and absorbs 22 keV gammas, only 88 keV line was used for calibration. The calibration spectrum is shown in Fig.\ref{fig3}.
The background measurements of Xe ($^{nat}$Xe) of natural composition were carried out before the sample no.~4 measurements. This sample of $^{nat}$Xe has been kept in underground conditions and thus had not been exposed to cosmic rays for about twenty years. The data of these measurements were used later for background subtraction in the dataset with radioactive isotope $^{127}$Xe.

\subsection{Production of $^{127}$Xe through channel
$^{128}$Xe~$(\mu,\gamma)^{127}$Xe}

Before these measurements, the Xe sample no.~4 has been kept in a special box with antineutron shield (1 mm cadmium and 20 cm polyethylene) in a room of "Carpet 2" installation \cite{r1} to expose the sample to cosmic ray’s muons .
The main reason of the antineutron shield was to prevent the production of $^{127}$Xe isotope by neutrons through $^{126}$Xe$(n,\gamma)^{127}$Xe channel.

\begin{figure}[pt]
\begin{center}
\includegraphics*[width=3.0 in,angle=0.]{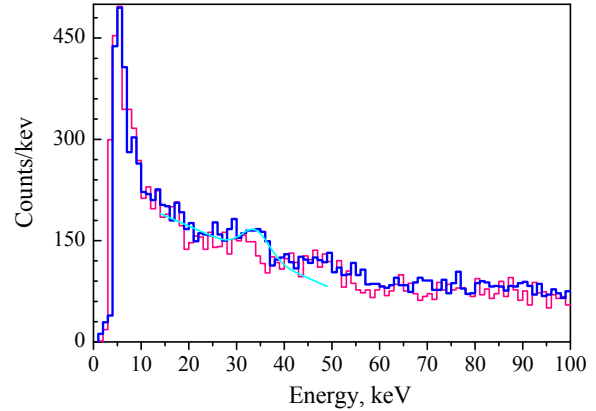}%
\caption{\label{fig4}
Amplitude spectra of LPC with xenon sample no.~4 - blue line, and with xenon sample of natural composition - red line,
and model calculation spectrum.
}
\end{center}
\end{figure}

The exposure time was 1968 hours. After the exposure sample no.~4 was taken to the underground laboratory and used to fill LPC. The working pressure of LPC was 1.3 atm. Measurements were carried out during 774 hours. Energy spectrum taken during this time is shown in Fig.\ref{fig4}.
The events from $^{127}$Xe decay were searched in the energy interval of $33.2 \pm 2\sigma$. The number of $^{127}$Xe events in the region of interest was determined by the difference between spectra of sample no.~4 and the background sample of $^{nat}$Xe. But the value obtained gives only the registered events. To find out the total number of $^{127}$Xe atoms that was at the beginning of measurements several renormalizations were done.

The first normalization was on the branching factor. $^{127}$Xe isotope decays to form $^{127}$I via capturing an electron from $K$-shell, while $^{127}$I nucleus stays in an excited state.
Therefore, full energy-release consists in contribution of characteristic photons and an Auge-electron appears during iodine $K$-shell filling; contribution of gamma rays and conversion electron appear after deexcitation of iodine nucleus.
Gamma rays have energy within $50\div370$ keV range.
The probability of gamma rays occurrence increases with energy and that of the conversion electrons decreases.
The efficiency of registration of gamma rays of high energy is very low. As a result, the main contribution to the region of interest is made by energy released in filling $K$-shall.
The fluorescence yield upon filling a single vacancy of the $K$ shell in iodine is 0.89\% \cite{x-ray}.
But in any case we need to take into account all decay channels in order to determine the amount of atoms at the beginning of measurements.

The second normalization is on the detection efficiency. This normalization allows one to evaluate the detection efficiency of the events with the energy sought for in this study and make recalculations for the full amount of events that occurred in the detector.
The detection efficiency for iodine characteristic photons is 0.26.
Using the obtained value of the LPC events and the law of radioactive decay we can get the equilibrium number of $^{127}$Xe atoms at the beginning of measurements.

The muon flux at the ground level where "Carpet2" installation is located is $(72 \pm 1.44)$~cm$^{-2}\cdot$h$^{-1}$ \cite{muon}.
Applying equation (\ref{eq1}) for the radioactive isotope recovery \cite{r7} we can calculate the cross-section of $^{127}$Xe isotope production from $^{128}$Xe by cosmic ray’s muons:
\begin{equation}
\label{eq1}
  \sigma  = \frac{{{N_o}}}{{\Phi  \cdot t \cdot \upsilon  \cdot {N_A}}},
\end{equation}
where $N_o=928 \pm 338$ is the number of $^{127}$Xe atoms that are in equilibrium at the beginning of measurements; $\Phi=(72 \pm 1.44)$~cm$^{-2}\cdot$h$^{-1}$  is the muon flux; $t=1968$ h is the exposure time of Xe sample; $\upsilon=0.3366$ mole is the amount of matter of $^{128}$Xe; $N_A=6.022 \cdot 10^{23}$ mol$^{-1}$ is the Avogadro's number. Putting the existing values into formula (\ref{eq1}), we get:
\begin{equation}
\sigma \leq 0.007 \cdot 10^{-24} \texttt{ cm}^2~\texttt{at 95\% C.L.}
\end{equation}

This rather conservative value is explained by the fact that the behavior of the Xe sample no.~4 and the Xe sample of natural composition in the region of interest is the same and the excess over background is minimal.

\subsection{Production of $^{127}$Xe through channel $^{126}$Xe$(n,\gamma)^{127}$Xe}

\begin{figure}
\begin{center}
\includegraphics*[width=3.0 in,angle=0.]{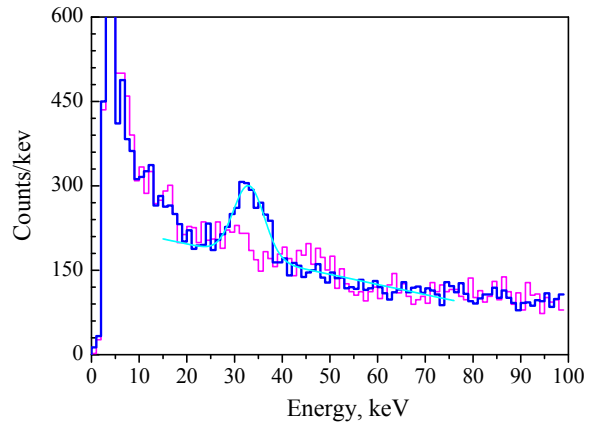}%
\caption{\label{fig5}
Amplitude spectra of  LPC with xenon sample no.~4 - blue line, with xenon sample of natural composition - red line,
and model calculation spectrum.
}
\end{center}
\end{figure}

For testing this channel of $^{127}$Xe production the bottle with the Xe sample under study has been placed in a separate room of the main building of BNO INR RAS. This room is located on the top floor of the building, where the flux of thermal neutrons produced by cosmic rays in the concrete blocks is maximal. The exposure time was 1272 hours. The neutron flux was measured by a special detector and was found to be $11.23 \pm 0.54$~cm$^{-2}\cdot$h$^{-1}$ \cite{neutro1, neutro2}. After the exposure the gas was taken to the underground low-background laboratory and used to fill LPC. The working pressure of LPC was 1.3 atm. Measurements were carried out during 1084 hours. The energy spectrum taken during this time is shown in Fig.\ref{fig5}.

The events from $^{127}$Xe decay were searched in the energy interval of $33.2 \pm 2\sigma$.
The technique of determining the total number of atoms $^{127}$Xe which was at the beginning of the measurements described above for the channel $^{128}$Xe$(\mu,\gamma)^{127}$Xe.

As before, the same formula (\ref{eq1}) was used. The following are the parameters for calculation of the cross section: $N_o=4922 \pm 490$ is the number of $^{127}$Xe atoms that are in equilibrium at the beginning of our measurements;
$\Phi=(11.23 \pm 0.54)$~cm$^{-2}\cdot$h$^{-1}$
is the neutron flux; $t=1272$~h is the exposure time of the Xe sample; $\upsilon=0.2089$ mole is the amount of matter of $^{126}$Xe.
The neutron capture cross section for $^{126}$Xe amounted to
\[
\sigma =(2.74 \pm 0.4)\cdot 10^{-24}\texttt{ cm}^2.
\]

In reference books the value of the cross section for $^{126}$Xe$(n,\gamma)^{127}$Xe reaction is equal to $\sigma =(3.5 \pm 0.8)\cdot 10^{-24}\texttt{ cm}^2$ \cite{r8}, that estimation is given for thermal neutrons with energy of $E_n=25.30$ meV. Our result is in good agreement, within the statistical error, with this reference value.

\section{Conclusions}

The measurements of $^{127}$Xe isotope production have been carried out.
In this work we have considered two channels of $^{127}$Xe production:
the neutron knock-out from $^{128}$Xe nucleus by cosmogenic muons and
the neutron capture by $^{126}$Xe nucleus.
For the first channel the upper limit of the cross section of $^{127}$Xe production has been obtained:
$\sigma \leq 0.007 \cdot 10^{-24}$~cm$^2$ at 95\% C.L.
For the second channel the value of the cross section was found to be equal to $\sigma =(2.74 \pm 0.4)\cdot 10^{-24}$~cm$^2$
which coincides well, within the statistical error, with the reference value.

\end{document}